%Paper: gr-qc/9502007
%From: Jorge Pullin <pullin@phys.psu.edu>
%Date: Wed, 1 Feb 1995 22:55:17 -0500 (EST)

\magnification=\magstep1
\vbadness=10000
\parskip=\baselineskip
\parindent=10pt
\centerline{\bf MATTERS OF GRAVITY}
\bigskip
\bigskip
\line{Number 5 \hfill Spring 1995}
\bigskip
\bigskip
\bigskip
\centerline{\bf Table of Contents}
\bigskip
\hbox to 6.5truein{Editorial and Correspondents {\dotfill} 2}
\bigskip
\hbox to 6.5truein{\bf Gravity news:\hfill}
\smallskip
\hbox to 6.5truein{LISA Recommended to ESA as Possible New Cornerstone
Mission, Peter Bender {\dotfill} 3}
\smallskip
\hbox to 6.5truein{LIGO Project News, Stan Whitcomb {\dotfill} 5}
\bigskip
\hbox to 6.5truein{\bf Research briefs:\hfill}
\smallskip
\hbox to 6.5truein{Some Recent Work in General Relativistic Astrophysics,
John Friedman{\dotfill} 7}
\smallskip
\hbox to 6.5truein{Pair Creation of Black Holes, Gary Horowitz{\dotfill} 10}
\smallskip
\hbox to 6.5truein{Conformal Field Equations and Global Properties of
Spacetimes, Bernd Schmidt{\dotfill} 12}
\bigskip
\hbox to 6.5truein{\bf Conference Reports:\hfill}
\smallskip
\hbox to 6.5truein{Aspen Workshop on Numerical Investigations of
Singularities in GR, Susan Scott{\dotfill} 15}
\smallskip
\hbox to 6.5truein{Second Annual Penn State Conference: Quantum Geometry,
Abhay Ashtekar {\dotfill} 17}
\smallskip
\hbox to 6.5truein{First Samos Meeting,
Spiros Cotsakis and Dieter Brill{\dotfill} 19}
\smallskip
\hbox to 6.5truein{Aspen Conference  on  Gravitational Waves and
Their Detection, Syd Meshkov{\dotfill} 20}
\smallskip
\bigskip
\bigskip
\bigskip
\bigskip
\leftline{\bf Editor:}

\medskip
\leftline{Jorge Pullin}
\smallskip
\leftline{Center for Gravitational Physics and Geometry}
\leftline{The Pennsylvania State University}
\leftline{University Park, PA 16802-6300}
\smallskip
\leftline{Fax: (814)863-9608}
\leftline{Phone (814)863-9597}
\leftline{Internet: pullin@phys.psu.edu}

\vfill
\eject

\centerline{\bf Editorial}

Well, I don't have much to say, just to to remind everyone
that suggestions and ideas for contributions are {\bf especially}
welcome. I also wish to thank the editors and contributors who made
this issue possible. The next newsletter is due September 1st.

If everything goes well this newsletter should be available in the
gr-qc Los Alamos bulletin board under number gr-qc/9502007. To
retrieve it send email to gr-qc@xxx.lanl.gov (or
gr-qc@babbage.sissa.it in Europe) with Subject: get 9502007 (number 2
is available as 9309003, number 3 as 9402002 and number 4 as
9409004). All issues
are available as postscript or TeX files in the WWW
http://vishnu.nirvana.phys.psu.edu

Or email me. Have fun.

\medskip
Jorge Pullin

\bigskip
\bigskip
\centerline{\bf Correspondents}
\medskip

\parskip=2pt
\item{1.} John Friedman and Kip Thorne: Relativistic Astrophysics,
\item{2.} Jim Hartle: Quantum Cosmology and Related Topics
\item{3.} Gary Horowitz: Interface with Mathematical High Energy Physics,
    including String Theory
\item{4.} Richard Isaacson: News from NSF
\item{5.} Richard Matzner: Numerical Relativity
\item{6.} Abhay Ashtekar and Ted Newman: Mathematical Relativity
\item{7.} Bernie Schutz: News From Europe
\item{8.} Lee Smolin: Quantum Gravity
\item{9.} Cliff Will: Confrontation of Theory with Experiment
\item{10.} Peter Bender: Space Experiments
\item{11.} Riley Newman: Laboratory Experiments
\item{12.} Peter Michelson: Resonant Mass Gravitational Wave Detectors
\item{13.} Stan Whitcomb: LIGO Project
\parskip=\baselineskip

\vfill
\eject

\centerline{\bf LISA Recommended to ESA as Possible New Cornerstone Mission}
\medskip
\centerline{Peter L. Bender, JILA, University of Colorado}
\centerline{pbender@jila.colorado.edu}
\bigskip

\parskip=7pt

     In May, 1993, two proposals for laser gravitational wave antennas
in space were submitted to the European Space Agency (ESA) as candidates
for the Third Medium-sized Mission (M3) under their Horizon 2000 program.
The Laser Interferometer Space Antenna (LISA) proposal was for laser
heterodyne measurements between four spacecraft in a cluster located well
behind the Earth, in orbit around the Sun.  The spacecraft in this
heliocentric antenna are located at the corners of an equilateral triangle
5 million km on a side, with two spacecraft at one of the corners, and the
plane of the triangle is tipped at 60 deg to the ecliptic.  The diameter
of the transmit/receive telescopes that send the laser beams between the
different spacecraft is 30 cm.  The frequency range of interest is from
somewhat below 0.1 millihertz to about 1 Hz.  Thus LISA and ground-based
gravitational wave detectors will complement each other by looking at
different frequency ranges.

     The other proposed mission, named SAGITTARIUS, has six spacecraft in
retrograde geocentric orbits with 0.6 million km radial distance.  Two
spacecraft are at each corner of an equilateral triangle with 1.0 million
km sides, which lies in the ecliptic.  The telescopes used have 15 cm
diameter.  Many other features of the two missions are the same.
SAGITTARIUS is intended to have lower cost because of smaller spacecraft
size and lower propulsion and telemetry requirements.  The sensitivity
for the SAGITTARIUS antenna is about a factor five worse than for LISA
over the frequency range of most interest.

     A common Assessment Study for the two missions [1] was carried out
by ESA during the period November, 1993, to April, 1994, with Dr. Yusuf
Jafry of the European Space Research and Technology Centre as the Study
Scientist.  The Chairs of the three Working Groups for the Study were as
follows:  Theory WG - B. Schutz, University of Wales, Cardiff;
Accelerometer WG - P. Touboul, Office National d'Etudes et de Recherches
Aerospatiales, France;  and Interferometry WG - J. Hough, University of
Glasgow.  The Study Team recommended the LISA mission, mainly because
of the factor five higher sensitivity and because the costs assigned by
ESA were not much lower for SAGITTARIUS.  However, LISA was not chosen
by ESA for further study as a candidate for the M3 mission because the
cost was roughly a factor two too high for an ESA medium-sized mission,
unless a major portion of the cost were to be provided by another
agency such as NASA.

     In October, 1993, a group led by Prof. Karsten Danzmann of the
Universitat Hannover and the Max-Planck-Institut fur Quantenoptik also
submitted the LISA concept for consideration by ESA as a new cornerstone
mission [2] under the proposed Horizon 2000 Plus program.  The concept
presented was the heliocentric mission with 5 million km antenna arm
lengths, but with six spacecraft instead of four.  Cornerstone missions
can be nearly a factor two more expensive than medium-sized missions,
and require ESA leadership.  They must have a scientifically valid
"core" that can be carried out by ESA alone, including launch and
operations, but might subsequently be enriched by new elements, to be
added through international collaboration(s).

     The Survey Committee appointed by ESA to make recommendations for
the Horizon 2000 Plus program has now recommended [3] the implementation
of three additional cornerstones (before the end of 2016) as follows:
	Cornerstone 5 (or 6) - Mission to Mercury
	Cornerstone 6 (or 5) - Interferometric Observatory
	Cornerstone 7        - Gravitational Wave Observatory.
The inclusion of the Fundamental Physics discipline (LISA) and the
recommended expanded Technological Activities will require a modest
increase in the funding of the ESA Scientific Programme beginning in
2001.  It is therefore proposed to augment the budget level by 5\%
each year for the years 2001 -2005.

     In the US, the NASA Astrophysics Division issued a Research
Announcement in September for proposals to study concepts for new
missions which could be flown after the year 2000.  If it is decided
to fund preliminary mission concept studies for participation in an
ESA gravitational wave mission, different possible forms of
collaboration with ESA will be considered.  However, the question of
future NASA participation in a LISA mission and the associated
scientific priority for such a mission have not yet been considered
by NASA or its advisory groups.

     The primary objective of the LISA mission is to search for and
study signals from sources involving massive black holes.  One
possible source is the coalescence of massive black hole binaries
formed by mergers of pre-galactic structures or galaxies.  The main
issue for predicting the number of such events is, at what stage in
structure formation are black holes with masses in the range of
roughly $10^4$ to $10^7$ solar mass likely to have formed?  The LISA
sensitivity is sufficient to detect such sources and to study them
in detail out to cosmological distances.  Another possible source
is compact stars orbiting around massive black holes in galactic
nuclei.

     Other objectives of the LISA mission are to observe signals
{}from galactic binaries and to search for a possible background of
gravitational radiation formed at early times.  LISA would certainly
observe signals from hundreds to thousands of neutron star binaries,
and would be able to determine their distribution throughout our
galaxy.  Comparable numbers of short period white dwarf binaries
probably will be seen also, and possibly enough to interfere with
the observations of some other types of sources.  Cataclysmic
variables and binary systems composed of a neutron star and a black
hole are also likely to be seen, as well as signals from some known
binaries.
\smallskip
\noindent {\bf References:}

\noindent [1]  LISA Study Team, "LISA:  Laser Interferometer Space Antenna for
gravitational wave measurements", in M3 Selection Process:  Presentation
of Assessment Study Results, 3 and 4 May, 1994, Paris (European Space
Agency SCI(94)9, May, 1994).

\noindent [2]  K. Danzmann et al, "LISA:  Laser Interferometer
Space Antenna for
Gravitational Wave Measurements (Cornerstone Mission Concept submitted
to ESA), Max Planck Institut f\"ur Quantenoptik, Garching (October, 1993).

\noindent [3]  European Space Agency, ESA SP-1180 (November, 1994).

\vfill
\eject

\parskip=5pt
{\centerline {\bf LIGO Project News}}
\medskip
{\centerline {Stan Whitcomb, Caltech}}
\centerline{stan@ligo.caltech.edu}
\bigskip

The Laser Interferometer Gravitational-wave Observatory (LIGO) project
was given a major boost in November when the National Science Board met
to consider LIGO and approved the new project plans.  This approval
included a revised construction estimate and subsequent funds for the
commissioning and initial operations.  The strong backing of the
National Science Foundation director and staff and the strengthening of
the project management were key elements in gaining the support of the
National Science Board.

The pace of work on the Project has increased substantially over the
past six months.  Site preparation is well underway at the Washington
LIGO site and should begin soon at the Louisiana site.  The rough
grading (the earthwork to level the foundation plane) has been
completed in Washington and the site will be allowed to settle while
the design of the foundations is finalized.  The site in Louisiana has
been purchased by Louisiana State University and leased to the NSF for
LIGO.  The Environmental Assessment has been completed; clearing of the
site will begin as soon as the final environmental approval is given.

A test of the beam tube design is now underway.  This is a test of the
design developed by our contractor (Chicago Bridge and Iron).  It
involves a full diameter section of beam tube approximately 40 m long,
fabricated with the techniques planned for the LIGO field
installation.  The key aspects of the design to be tested are the
leak-tightness of the welds and the outgassing of the fabricated tube.
The tube is now under vacuum and a bake-out of the tube (140 C for 30
days) is planned to start in early February.

The Ralph M. Parsons Company was selected as the Achitect-Engineer for
the LIGO facilities.  They will design the building for the two LIGO
sites, including the foundations and covers for the beam tubes.  They
will also take responsibility for the site planning and eventually
provide management oversight for the actual construction.  Parsons was
selected after a nationwide solicitation and a very rigorous
competition.

The final major facilities design contract is for the vacuum system,
including the chambers, pumping system, and vacuum instrumentation, but
excluding the beam tubes.  Proposals for this contract are due in
February and we hope to have the design work underway before summer.

The major highlight from the R\&D program is another improvement in the
sensitivity of the LIGO 40 m interferometer.  New test masses were
installed in the interferometer.  These new test masses are of a
monolithic design with the mirror surface an integral part of the test
mass; the earlier test masses had a compound construction with mirrors
optically contacted onto a fused silica body.  The compound design
appears to have been a source of noise.  The interferometer now has a
peak sensitivity  (near 450 Hz) of $2.5 \times 10^{-19}
\rm{m}/\rm{Hz}^{1/2}$, expressed as an equivalent differential arm
length.

The LIGO Project is now operating a World Wide Web server which will
provide access to general information about LIGO, latest news and
results, preprints and technical reports, and other relevant
information.  You can access our home page
(our URL is ``http://www.ligo.caltech.edu'') using Mosaic or another
WWW browser.  We expect that this will become one of the principal
channels of communication with the interested scientific community.

\vfil
\eject

\centerline{\bf Some Recent Work in General Relativistic Astrophysics}
\medskip
\centerline{John Friedman, University of Wisconsin-Milwaukee}
\centerline{friedman@thales.phys.uwm.edu}
\bigskip
\bigskip
\medskip

This will be a short summary highlighting aspects of work on three
different topics:  numerical models of coalescing neutron stars; an
update on work that sharply restricts the role that nonaxisymmetric
instability could play in rotating neutron stars; and an announcement
of a public-domain code and some other accurate recent codes for
modeling rotating relativistic stars.

$\bullet$ Several groups have begun detailed studies of the late stages of
coalescing neutron-star binaries, elucidating much more of physics
immediately preceding merger.  An orbital instability known from work
on Newtonian ellipsoids appears to dominate the final part of the
inspiral, and it can be found analytically as well as by the numerical
hydrodynamics that has been implemented. For a $r^{-1}$ potential,
circular orbits are, of course, always stable.   In general relativity
an effective potential that rises more quickly leads to unstable
circular orbits within $r=6 M$ for small particles orbiting larger
masses; one expected that the same instability for a system of with
components of roughly equal mass would limit the smooth inspiral of two
black holes or possibly even two neutron stars if the equation of state
is soft enough that unstable orbits lie outside the stars' surfaces.
But there is a surprise.  The {\sl Newtonian} gravitational force
between two stars in a binary system already has a tidal contribution
that rises more quickly than $1/r^2$, fast enough that the effective
potential can have a maximum.  For  incompressible ellipsoids in a
binary system, this leads to an unstable orbit before the ellipsoids
coalesce, and Rasio and Shapiro and coworkers have shown that the same
dynamical instability arises for models as compressible as neutron
stars.

The published studies of binary coalescence of neutron stars have used
Newtonian gravity, generally corrected by a quadrupole gravitational
radiation-reaction term.  Shibata, Nakamura and Oohara used an Eulerian
code to look at synchronously rotating binaries and at binaries with
nonrotating stars.  The viscosity of neutron stars is probably too
small to synchronize their spins in the time it takes them to spiral
together (Kochanek; Bildsten and Cutler), and three groups (Davies,
Benz, Piran and Thielmann; Colpi, Rasio, Shapiro, and Teukolsky; Zhuge,
Centrella and McMillan) have used smoothed particle hydrodynamics to
study binaries without synchronized spins.  For stars of $1.4 M_\odot$
and radius 10 km, orbital instability at a 30 km separation leads to
rapid merging, with 20\% of the total mass ejected from
the central region to form dramatic, if short-lived, spiral arms.
Now neutron stars near the maximum  {\sl and minimum} mass configurations
are unstable to radial oscillations, and Clarke, Eardley and Blinnikov had
suggested that tidal stripping of matter in coalescence could lead to an
explosion of the less massive member of a coalescing binary system when
it nears its minimum mass.  Orbital instability and merger is the picture
given by the current simulations, but Colpi and Rasio suggest that the
spiral arms quickly fragment into lumps smaller than the minimum mass
which then explode.

$\bullet$
Millisecond pulsars are being detected at an accelerating rate.  A dozen
are now known with periods less than 3 ms, although the shortest known
period (1.6 ms) is still that of the first fast pulsar.  The upper
limit on the angular velocity of a neutron star is sensitive to the
equation of state of matter above nuclear density, and there is a
reasonable chance that within the next decade we will have confidence
in an observational value of that limit.  For neutron stars with
sufficiently weak magnetic fields, the upper limit on rotation is set
by gravity, but there has been a question of whether a nonaxisymmetric
instability driven by gravitational radiation will limit the rotation
before the equator of the star rotates at the Kepler frequency.
Lindblom and Mendell have completed an analysis of the damping of
normal modes of neutron stars by an effective viscosity arising from a
superfluid dissipation mechanism called `mutual friction'.

In rotating neutron-star matter mutual friction is caused by the
scattering of electrons off the cores of the neutron vortices. This
scattering is greatly enhanced by nuclear interactions between the
neutrons and protons that induce proton supercurrents and hence strong
magnetic fields within the neutron vortices.  Lindblom and Mendell have
spent several years on a careful treatment of the superfluid interior
of neutron stars, and the application of their formalism to the
nonaxisymmetric instability appears to be unambiguous and striking:
Mutual friction completely suppresses the gravitational-radiation
instability in all neutron stars cooler than the superfluid-transition
temperature.  They take pains to enumerate caveats to the conclusion,
but the work seems to leave little doubt that the nonaxisymmetric
instability plays no role in limiting the rotation of old neutron stars
spun-up by accretion.  The paragraph itself needs a caveat:  Even with
a weak magnetic field, and without the gravitational instability, a
medium-to-soft equation of state might mean that a neutron star will be
spun up by accretion to a limit shy of the Kepler frequency, because
the innermost stable circular orbit (and hence the inner edge of the
accretion disk) can lie outside the star.

If rapidly rotating neutron stars form from the accretion-induced
collapse of white dwarfs, there may still be a time during the cooling,
at temperatures between $1\times 10^{10}$ K and the superfluid
transition temperature of about $10^9$ K during which the
nonaxisymmetric instability could play a role.  But earlier work of
Ipser and Lindblom (see Lindblom 1995 and references therein, and
Hashimoto et.  al. for updated and corrected calculations) together
with the damping of the instability at $10^9$ K means that the
deviation from the Kepler frequency would be small.

$\bullet$
Responding to rapidly growing class of fast pulsars, a number of
different groups have written codes to construct models of rapidly
rotating neutron stars. (The methods in use are due to Butterworth and
Ipser (BI); to Komatsu, Eriguchi and Hachisu (KEH), who use the
Butterworth-Ipser set of equations but a somewhat different algorithm;
Cook, Shapiro and Teukolsky with a modification of KEH; Neugebauer and
Herlt, a finite-element method; and Salgado et al, with a alternative
algorithm and an alternative set of equations of state.)  An ongoing
comparison of substantially different codes so far shows agreement
limited only by grid size and accuracy that is easily better than 1% in
all quantities (vastly better than the uncertainty in the equation of
state from which the models are constructed).  A public-domain code,
written by Stergioulas, is now available, and may make models of
rotating stars nearly  as accessible as spherical models.  The
Stergioulas code is automated to construct sequences of constant
angular momentum and constant baryon number and to locate
configurations with maximum angular velocity.   It implements the KEH
method as modified by Cook et al., and agrees to better than 0.1\% on
with a recent code accurate code constructed by Gourgoulhon et al. and
based on a substantially different method and a different choice of
independent field equations.

As Cook et al.  have found, there is often a slight difference between
the models of maximum $M$ and $\Omega$ for models of neutron stars
corresponding to a proposed equation of state.   In fact there are two
classes of equations of state, depending on whether the model with
maximum mass among all equilibrium configurations is unstable
(Stergioulas and Friedman).  If it is, the {\it stable} models with
maximum mass, baryon mass, angular velocity, and angular momentum
coincide.  Otherwise they are all distinct, although in general close
enough that a dense set of models is needed to resolve them.

\bigskip
\parskip=4pt

\noindent
{\bf References} (Only one reference by the same set of authors is listed.):

\noindent Butterworth, E. M., \& Ipser, J. R. ApJ, 204, 200 (1976).

\noindent Colpi, M. and Rasio, F., ``Explosions of neutron star fragments
ejected
during binary coalescence,'' preprint (1994).

\noindent Cook, G. B., Shapiro, S. L., \& Teukolsky, S. A.  ApJ, 424, 823
(1994).

\noindent Bildsten, L. \& Cutler, C. ApJ, 400, 175 (1992).

\noindent Davies, M, Benz, W., Piran, T., \& Thielmann, F., Ap J 431, 742
(1994).

\noindent Hashimoto, M., Oyamatsu, K., \& Eriguchi, Y. ``Upper limit of the
angular velocity of neutron stars'', preprint 1994.

\noindent Kochanek, C. Ap J 398, 234 (1992).

\noindent Komatsu, H., Eriguchi, Y., \& Hachisu, I.  MNRAS, 237, 355 (1988).

\noindent Lai, D., Rasio, F. \& shapiro, S. Ap. J. Suppl. 88, 205 (1993).

\noindent Lindblom, L. ApJ 438, 265 (1995).

\noindent Neugebauer, G., \& Herlt, E. Class. Quant. Grav., 1, 695 (1984).

\noindent Salgado, M., Bonazzola, S., Gourgoulhon, E., and Haensel, P.,
Astron. and Ap., in press.

\noindent Stergioulas, N. and Friedman, J.L., ApJ in press.

\noindent Lindblom, L, \& Mendell, G. Ap. J. in press.

\noindent Shibata, M. Nakamura, T. and Oohara, K. Prog. Theor.
Phys 88, 1079 (1992).

\noindent Rasio, F. and Shapiro, S. Ap J 432, 242 (1994).

\noindent Zhuge, X., Centrella, J., and McMillan, S. L. W.  ``Gravitational
radiation from coalescing binary neutron stars,'' preprint 1994.

\parskip=7pt
\vfill
\eject

\centerline{\bf Pair Creation of Black Holes}
\medskip
\centerline{Gary T. Horowitz, UCSB}
\centerline{gary@cosmic2.physics.ucsb.edu}
\bigskip
\bigskip

Hawking's prediction of thermal radiation from a black hole remains
one of the most important results to emerge from the union of quantum
mechanics and gravity. There are good arguments that this prediction
is independent of unknown Planck scale physics for black holes much
larger than the Planck mass. However there is another prediction that
can be made which combines quantum mechanics and gravity, and appears
to be independent of Planck scale physics. This is the fact that
charged black holes will be pair created in a background electric or
magnetic field. This process is the direct analog of the creation of
electron-positron pairs or monopole-antimonopole pairs which have been
studied previously.  Over the past few years, the pair creation of
black holes has attracted considerable attention.  A key motivation is
that, in addition to being an apparently unambiguous consequence of
quantum gravity, this process is likely to shed light on the nature of
black hole entropy.  Since a black hole should be equally likely to be
pair created in any of its available states, by comparing the rate for
black hole creation to the rate of e.g. monopole creation, one can
effectively count the number of internal states of a black hole.

In direct analogy to tunneling phenomena in ordinary field theory,
black hole pair creation can be described in a semi-classical
approximation using an instanton. As Gibbons first pointed out
[1] , the exact instanton describing this pair creation event can
be constructed by analytically continuing a solution found by Ernst in
the mid 1970's.  This instanton was first studied by Garfinkle and
Strominger [2] who argued that regularity of the instanton fixed the
charge to mass ratio of the black holes so that they were always
nonextremal. They also found that the black holes were created with
their horizons identified, so that they formed a wormhole in space.
The rate of pair creation (in the semi-classical approximation) is
directly related to the action of the instanton. Garfinkle and
Strominger computed this action in the limit of weak fields (i.e. $qB
<<1$) and found that it agreed with the rate of pair creating
monopoles with the same mass and charge. Of course, the most likely
black holes to be produced have the Planck mass, and for these, the
semiclassical approximation breaks down. However, this approximation
should be valid for the pair creation of larger black holes, which is
predicted to occur with a nonzero (although small) probability.

The next step forward was taken by Garfinkle, Giddings, and Strominger
[3] who computed the instanton action exactly and found that it was
smaller than the the corresponding monopole case by precisely a factor
of the Bekenstein-Hawking entropy $S=A/4$. In other words, the pair
creation of black holes is enhanced over the pair creation of
monopoles by a factor of $e^S$, exactly what one would expect if black
holes had $e^S$ internal states.

This simple state of affairs did not last long. Dowker et. al. [4]
soon found that there was another regular Ernst instanton which
described the pair creation of extremal black holes. When they
computed its action, they found that the extremal pair creation rate
was NOT enhanced over that of monopoles. This puzzle was clarified in
a recent paper by Hawking et. al. [5] who gave independent arguments
that the entropy of an extreme Reissner-Nordstrom black hole is
zero. The discontinuity between nonextreme black holes (which have
$S=A/4$) and extreme black holes was related to the change in topology
of the euclidean solutions. Thus pair creation arguments provide
evidence that black hole entropy is indeed related to the number of
internal states of a black hole.

An important qualification needs to be made at this point. The above
statements about the rate of pair creation are based solely on the
leading order semiclassical approximation. Higher order terms
(e.g. fluctuations about the instanton) have not been included and may
be large (especially in the extremal case).  It is important to
understand these corrections because for ordinary field theory
instantons, the density of states factor comes from the one loop
fluctuations. For black hole pair creation, it appears to be present
in the instanton action. Although this is unusual from the field
theory standpoint, it is not unprecedented. It is well known that one
can derive the Bekenstein-Hawking entropy for a single black hole from
just the black hole instanton, without including the fluctuations.

In related work, Dowker et. al. [6] found a generalization of the
Ernst instanton which includes an arbitrary coupling to a dilaton and
discussed the pair creation of charged dilatonic black holes. For a
particular value of the dilaton coupling, the extremal limit describes
the pair creation of Kaluza-Klein monopoles [4]. Hawking
et. al. [5] also pointed out that if black holes can be pair
created, it should be possible for them to annihilate, and discussed
some consequences of black hole annihilation.

Before the experimentalists get too hopeful, I should perhaps point
out that it would take a magnetic field of approximately $10^{50}$
gauss to have a reasonable probability of pair creating magnetically
charged black holes.  Unfortunately, electric fields are even less
promising, since they would decay through electron-positron pair
creation long before black holes were produced.

\bigskip
\noindent {\bf References}

\noindent [1] G. Gibbons in {\it Fields and Geometry} ed. A. Jadczyk
(World Scientific, 1986).

\noindent [2] D. Garfinkle and A. Strominger,  Phys. Lett., B256, 146, 1991.

\noindent [3] D. Garfinkle, S. Giddings and A. Strominger,
Phys. Rev., D49, 958, 1994.

\noindent [4] F. Dowker, J. Gauntlett, S. Giddings and G. Horowitz,
Phys. Rev., D50, 2662, 1994.

\noindent [5] S. Hawking, G. Horowitz, and S. Ross, `Entropy, Area, and
Black Hole Pairs", submitted to Phys. Rev., gr-qc/9409013.

\noindent [6] H.F. Dowker, J.P. Gauntlett, D.A. Kastor and J. Traschen,
Phys. Rev.,  D49, 2909, 1994.
\parskip=7pt

\vfill
\eject

\centerline{\bf The Conformal Field Equations and Global
Properties of Spacetimes }
\vskip1cm
\centerline{Bernd G. Schmidt}
\centerline{ Max-Planck-Institut f{\"u}r Astrophysik, Garching bei M\"unchen}
\centerline{bgs@mpa-garching.mpg.de}
\bigskip
\bigskip

H. Friedrich's method of ``regular conformal field equations" has
shown to be very useful to establish existence and asymptotic
properties of solutions of Einstein's field equations

$$
\tilde R_{ab}=\Lambda\tilde g_{ab}\ . \eqno(1)
$$

	I shall try to outline this method and some results achieved by it.
 References
can be found in [1].

Motivated by Penrose's treatment of null infinity via a conformal rescaling of
 the spacetime
 metric $\tilde g_{ab}$ the aim is to use the ``unphysical metric"
$$
g_{ab}=\Omega^{2}\tilde g_{ab}\ \eqno(2)
$$
together with the conformal factor $\Omega$ as the unknown fields.
Rewriting of (1) in terms
of the unphysical metric and $\Omega$ leads to the equation
$$
\tilde R_{ab}=R_{ab}+2\Omega^{-1}\nabla_a\nabla_b\Omega
+g_{ab}g^{cd}
\left(\Omega^{-1}\nabla_c\nabla_d\Omega-3\Omega^{-2}
\nabla_c\Omega\nabla_d\Omega\right)
=\Lambda\Omega^{-2}g_{ab}\ .\eqno(3)
$$
This equation is singular for $\Omega=0$, that is precisely at those points
(at infinity) where
we want to understand its consequences.

The ``regular conformal field equations" are a system of equations
(for the conformal factor, the rescaled metric, the non--physical
Ricci tensor and the rescaled Weyl tensor
$d^a{}_{bcd}:=\Omega^{-1}C^a{}_{bcd}$) which are equivalent to (3) and
regular in the sense that no factor of $\Omega^{-1}$ occurs in the
equations and that $\Omega$ does not appear in the principle part of
the differential operator associated with the equations (the terms
with the highest derivatives). The Bianchi identities are part of the
equations.

These equations are not only regular, but can furthermore be split
into a system of symmetric hyperbolic evolution equations and
constraints which are compatible with the evolution. This allows to
study various Cauchy and characteristic initial value problems with an
initial hypersurface on which $\Omega$ vanishes, hence to prescribe
data at points which are at infinity of the physical spacetime.

It is worthwhile to note that the evolution equations are in a certain
sense hyperbolic in any coordinate system.

I shall now consider the three cases $\Lambda>0$, $\Lambda<0$, and
$\Lambda=0$ and describe some of the results obtained.

The de Sitter solution has a positive cosmological constant and is
geodesically complete. It can be conformally embedded into the
Einstein universe and its boundary consists of two spacelike
hypersurfaces $scri^+$ and $scri^-$ on which $\Omega$
vanishes. Therefore this embedding defines a solution of the conformal
field equations. One can now analyze the Cauchy problem to find out
which data can be given on a spacelike hypersurface on which
$\Omega=0,d\Omega\neq 0$. The result is that a positive definite
metric is freely specifiable together with the electric part of the
rescaled conformal Weyl tensor. De Sitter data determine uniquely the
de Sitter solution. If we take data sufficiently near these data,
general theorems on the stability of solutions of symmetric hyperbolic
systems on compact domains imply that these solutions exist on a
sufficiently large domain and reach a second hypersurface on which
$\Omega$ vanishes. Translating this result to physical spacetime, we
have constructed a solution which is geodesically complete and
asymptotically de Sitter in the past as well as in the future.

Suppose we would like to prove the same theorem working in physical
spacetime.  Changing the data from de Sitter data on some Cauchy
surface would by general theorems only give a solution on some compact
part of the de Sitter spacetime.  To obtain a solution which is
geodesically complete special estimates would be needed, and no
general method is known to obtain such estimates. Thanks to the
conformal equations no such estimates are needed.  ``The geometry is
worked out of the equations".

Recently H. Friedrich, [1], considered the anti--de Sitter spacetime
($\Lambda<0$). Again it can be conformally embedded into the Einstein
universe. Its boundary is timelike with the topology $S^2\times R$. To
prove existence of solutions which are asymptotically anti--de Sitter
Friedrich solved a boundary initial value problem for the conformal
field equations with a timelike boundary. This is the first general
initial boundary value problem in the context of Einstein's equations
which found complete treatment. Translating again to physical
spacetime the solution behaves asymptotically anti--de Sitter near
$scri^+$ which is timelike.

For de Sitter and anti--de Sitter space the boundaries of the
conformal imbeddings into the Einstein universe are smooth
hypersurfaces, and it is possible to pose regular initial or boundary
value problems. This is different for the conformal embedding of
Minkowski space into the Einstein universe. Besides the smooth null
hypersurfaces at infinity, $scri^+$ and $scri^-$, there are the
vertices of these null cones, $I^0$ and $I^\pm$ where $\Omega$
vanishes. Examples like the Schwarzschild spacetime show that in
general the conformal structure will be singular at spacelike infinity
$I^0$.

Postponing the problem of spacelike infinity the hyperboloidal Cauchy
problem was studied. Prescribing almost Minkowski data on a
hypersurface intersecting $scri^+$, H. Friedrich showed the existence
of solutions which are geodesically future--complete and have a
regular point $I^+$ as future timelike infinity. So far no such a
result has been obtained by working with the equations in spacetime.

Numerical investigations of a hyperboloidal Cauchy problem by
P.H\"ubner, [2], using conformal field equations coupled to a scalar
field in the case of spherical symmetry, demonstrated that one can
calculate numerically global properties (horizon, $scri^+$,
singularities) of the spacetime on a compact grid with regular
equations.

To treat the usual Cauchy problem by conformal techniques one has to
analyze the singularity at spacelike infinity enforced by positive
ADM--mass.  Investigations are under way to decide which class of data
will evolve into spacetimes with smooth $scri^+$, and will finally
resolve the general structure of spatial infinity.  \vskip 1cm
\item{[1]}{Helmut Friedrich; Einstein Equations and Conformal Structures:
Existence of Anti--de
Sitter Space--Times; MPA 808 June 1994; to appear in Journal of
Geometry and Physics}

\item{[2]}{Peter H\"ubner; }{A Method for Calculating the Structure
of (Singular) Spacetimes in
the Large; GR--QC--940929}

\vfill
\eject

\centerline{\bf Aspen workshop on numerical investigations of
singularities in general relativity}
\medskip
\centerline{Susan Scott, Australian National University}
\centerline{Susan.Scott@anu.edu.au}
\bigskip
\bigskip

This workshop was organized by Jorge Pullin, Bernd Schmidt, Susan
Scott and took place in the Aspen Center for Physics from August 22 to
September 11 1994.  Activity at the workshop was mainly concentrated
in the following three areas:

\noindent $\bullet$ The new type of singularity structure discovered
by Choptuik in
gravitational collapse.

\noindent $\bullet$ The status of theoretical developments in singularity
theory.

\noindent $\bullet$ The interpretation of numerically observed singular
space-times.

The first week of the workshop focussed mainly on the study of the
phenomena discovered by Choptuik. In brief, Choptuik considers the
collapse of a spherically symmetric scalar field. The final outcome can
either be a black hole or empty space; in both cases the scalar field
radiates to infinity. Choptuik concentrates on the space-times in the
boundary between the two behaviors cited above. He finds that the mass
of the final black hole created has a universal law as a function of the
initial data that closely resembles the power laws observed in critical
phenomena in statistical mechanics. Choptuik also observes that these
space-times present a unique universal pattern of oscillations with a
discrete self-symmetry.

The workshop opened with a talk by Charles Evans who presented a model
similar to that of Choptuik, but where the scalar field was replaced by a
perfect fluid. He has been able to find exactly a critical self similar
solution and expects to be able to find the critical exponent by using
perturbation theory. Richard Price talked about a very different approach
in which analytic approximations were studied taking Choptuik's data as
``experimental data''. Douglas Eardley presented yet another model, with
a complex scalar field, where again an exact critical solution can be
found. This led to a lot of discussion concerning the possible
relationship with Evans' model. In the background of all these
discussions were several connections with the empirical observations of
Price on Choptuik's data.

Carsten Gundlach presented a completely different approach where the
discrete self-similarity observed by Choptuik is taken as exact and
the equations are integrated as a boundary-value problem within a
single self-similar region. The expectation is that by requiring
regularity at the origin and at a ``sonic horizon'', the critical
Choptuik solution will appear as a unique solution. John Stewart
described an approach to the Choptuik space-time based on the
Newman-Penrose formalism and the characteristic initial value
problem. He also described some preliminary calculations that show
that an approach similar to the one Price presented in his lecture can
be taken in the Newman-Penrose language. It was an example of a piece
of research directly motivated by the communication fostered by the
workshop.

The phenomena discovered by Choptuik have been the source of a lot of
discussion and excitement in the general relativity community worldwide
in recent times. This was the first workshop partly devoted to this
topic. It led to much discussion and comparison of notes among a number
of the experts who are trying to understand this problem.

There was a series of three talks given by Chris Clarke, Bernd Schmidt
and Susan Scott which gave an overview of the theoretical developments in
singularity theory for relativity during the past three decades. These
included the main topics of the classification of singularities, boundary
constructions for space-time and the possibility of performing extensions
of space-times. This series provided a very useful and timely background
for many of the numerical relativists present.

A number of interesting theoretical investigations were either started or
undertaken during the workshop. Chris Clarke and Susan Scott developed
four different ways of topologising a manifold together with its abstract
boundary --- the abstract boundary is a boundary construction recently
developed by Scott and Szekeres which is applicable to any $n$-dimensional
manifold. Steven Harris and Susan Scott considered how causal structure
could be included in the abstract boundary construction for space-time.
Steven Harris also made some progress in his attempt to prove that
compactness is a property of the Busemann boundary of a general
Riemannian manifold.

Numerical investigations by Beverly Berger, David Garfinkle and Vince
Moncrief of vacuum cosmological models with compact space sections with
two Killing vectors reveal a surprisingly simple singularity structure.
It is ``velocity-dominated'' which implies, in particular, that the
asymptotic geometry is known analytically. A code for the case with just
one Killing vector is about to bring results in the coming months. If it
turns out the singularities are also velocity-dominated in this quite
general case, numerical relativity has produced a challenge for
mathematicians to prove the existence and understand the basic mechanisms
that give rise to these singularities. For this topic the workshop was
again a unique opportunity, since it brought together the people doing
the actual numerical work with several mathematical experts on
singularity theory.

Finally, there was a series of talks dealing with subjects somewhat
related to singularities. Jorge Pullin, Jim Wilson and Jeff Winicour
spoke about collisions of stars and black holes from different points of
view, and Hans-Peter Nollert described his recent work that tends to
suggest that new normal modes exist in the formation and distortion of
black holes.

\vfil
\eject
\centerline{\bf Second Annual Penn State Conference: Quantum Geometry}
\medskip
\centerline{Abhay Ashtekar, Penn State}
\centerline{ashtekar@phys.psu.edu}
\bigskip
\bigskip
\medskip

This series of conferences was inaugurated in '93 and the first
meeting, held in October of that year, was devoted to numerical
relativity. The second conference was held during the first three days
of September '94 and its theme was quite different. It was motivated
by the fact that, over the last few years, several distinct approaches
have been developed to explore the nature of quantum geometry. Some
come from advances in quantum gravity, some from the mathematical
developments on low dimensional manifolds and some from recent
advances in computational physics. The aim of this conference was to
bring together experts in these different areas to discuss recent
results and to enhance the dialog between these three communities.  It
was a ``discussion conference'' with approximately 80 participants
{}from US, Canada, Mexico and Europe.

There were relatively few, long talks with ample time for questions,
discussion and long comments. The exchanges that followed the main
talks were generally lively and led to a deeper understanding of the
state of the art. Indeed to non-experts, these discussions were
sometimes more illuminating than the talks themselves. There was a
healthy mix of physicists and mathematicians although, unfortunately,
the computational physicists were under-represented.

The conference program can be divided into five broad categories. The
first of these dealt with recent mathematical developments and their
applications. It included talks by John Baez on ``Higher dimensional
algebras and topological quantum field theories,'' John Barrett on
``Quantum gravity as a topological field theory,'', Roger Brooks on
``Quantum gravity and equivariant cohomology'' and Daniel Kastler on
``Non-commutative geometry and its applications to the standard model
of particle physics''.  Baez's beautiful synthesis of the interplay
between algebras and TQFTs in two, three and four dimensions is
briefly summarized in the introduction to his contribution to the
Marcel Grossmann Proceedings and is available on the
network. Similarly, the tantalizing ideas on ``geometrizing'' the
Higgs mechanism using Alain Connes framework are discussed in
Kastler's Luminy lecture notes, also available electronically.  The
second set of talks concerned computational physics. Here, Jan Ambjorn
provided a broad review entitled ``Computer simulations of quantum
gravity: A viable approach?'' and Bernd Br\"ugmann discussed subtle
issues related to the choice of measures in his talk ``Dynamical
triangulations of four dimensional quantum gravity.'' The third set of
talks discussed the insights that have been gained into the nature of
quantum geometry through two approaches to quantum gravity, string
theory and non-perturbative quantum general relativity. Jorge Pullin
presented a general review in ``Recent mathematical developments
in non-perturbative quantum gravity;'' Carlo Rovelli presented recent
results on the spectra of geometric operators and on quantum dynamics
in a talk entitled ``In search of topological Feynman rules for
quantum gravity'' and Brian Greene, in his talk ``Mirror symmetry and
space-time topology change,'' explained how differential geometry is
`transcended' in string theory.

The last two sets of talks was related to quantum geometry only
indirectly. However, the areas they covered are fertile and just of
the sort that may lead to new insights into this subject. The first
dealt with the issue black hole entropy and the number of
microstates. Ted Jacobson summarized the new developments and shared
his understanding of the subject as a whole in his talks ``Black hole
entropy and vacuum fluctuations'' and Steve Carlip, in his talk
``Statistical mechanics of the 3-d black hole,'' presented some very
tantalizing ideas about the degrees of freedom residing on the black
hole horizon. The last set of talks was devoted to mathematical and
conceptual issues in classical and quantum general relativity and
Yang-Mills theory. Viqar Husain discussed his recent work on
``Self-dual gravity and the chiral model;'' Charles Torre provided an
elegant summary of his work in ``Hidden symmetries, observables and
symplectic structures for the Einstein equations: just say NO;''
Ranjeet Tate discussed the ``Singularities in quantum minisuperspace
models;'' Arlen Anderson sketched his recent ideas on ``The issue of
time in quantum gravity'' and Ingemar Bengtsson pointed out subtleties
in quantizing constrained systems in his talk ``Yang-Mills on a circle
--Non compact gauge groups.'' There was also a thought provoking after
dinner talk by Roger Penrose entitled ``Shadows of the mind'' which
provided a preview of his book which has since been published. While
this was a general talk, it was most fitting for the subject of the
conference because of Penrose's long held view that it is some aspect
of quantum gravity that would be responsible for the non-computability
inherent in the functioning of the human mind.

This series of conferences does not normally publish proceedings.
However, detailed reviews based on most of the main talks at this
specific conference (together with a few invited articles) will be
published as a special issue of the Journal of Mathematical Physics
in the fall of 1995. The issue will be entitled ``Quantum geometry and
diffeomorphism invariant quantum field theory'' and will be edited by
Lee Smolin and Carlo Rovelli.

\vfil
\eject

\centerline{\bf First Samos Meeting on Cosmology, Geometry and Relativity}
\bigskip
\centerline{Spiros Cotsakis, University of the Aegean and Dieter Brill,
University of Maryland}
\centerline{skot@pythagoras.aegean.ariadne-t.gr, brill@umdhep.umd.edu }
\medskip
\parskip=5pt
Ten years ago a new university was founded in the Aegean
Archipelago. In the 1990's general relativity research was begun in
its Department of Mathematics, which is located on the island of Samos
in the east Aegean sea.  Samos will be known to many relativists as
the birthplace of Pythagoras; this year it has firmly established its
place on every relativist's world map through an international
relativity conference that promises to be the first in a series. This
year's meeting was held at Karlovassi on  September 5-7 1994,
preceded by the 2nd Summer School on Analysis, Geometry and
Mathematical Physics in the Department of Mathematics (Organizers:
M. Anoussis, S. Cotsakis and N. Hadjisavvas), which focused
on Geometry and Cosmology. It was intended for Greek undergraduate and
graduate students of mathematics, physics and related fields.  The
plan is to have these Schools every year as a forum and bridge for
entering research in these fields.

The First Samos Meeting on Cosmology, Geometry and Relativity
was organized by S.~Cotsakis (Aegean) and G.~W.~Gibbons
(Cambridge). It was attended by about 30 participants, half from Greece and
the remainder from countries as diverse as Europe, India, South Africa, and
the US. The aim of the meeting was to show the close interplay between
rigorous mathematics and its use in relativity and cosmology. Rather than
featuring a great variety of topics, a smaller number of invited speakers
were given the opportunity to develop their subjects in depth and detail by
allotting two hours for each lecturer.

The main speakers and their topics were: \smallskip
\settabs \+in&name of speaker here &title \cr
\+&Y.~Choquet-Bruhat &Global existence for ultrarelativistic Yang-Mills
fluids\cr
\+&D.~Brill &Black hole collisions, analytic continuation and cosmic
censorship\cr
\+&G.~W.~Gibbons &Gravitating solitons\cr
\+&D.~Christodoulou &Relativistic fluids and gravitational collapse\cr
\+&R.~Beig &The Einstein vacuum constraints and trapped surfaces\cr
\+&V.~Moncrief &Hamiltonian reduction of Einstein's equations.\cr

The intensive pace of the lectures and contributed papers was relieved by
social hours and a banquet. The fine weather allowed these to be held in the
open air, symbolizing to us the open and friendly nature of the meeting
and its location. We also had welcome opportunity to appreciate the beauty
and long cultural tradition of this island. In accordance with local
customs, the lunch breaks lasted four hours, allowing us to explore the town
of Karlovassi, its beaches, and the nearby countryside. The conference ended
with an afternoon excursion to the principal archeological sites, such as
the Vathi Museum and the Temple of Hera, and a tour around the island
with lunch in a traditional Greek tavern by the beach.

During the Meeting it became clear that all participants were very positive
about the idea of a 2nd Samos Meeting on Geometry, Cosmology and Relativity.
Accordingly it was decided to have the second meeting sometime around the
summer of 1996. As plans develop details will be posted on  MacCallum's
gr-list. The Proceedings of the conference will be published in the
Springer series  {\it Lecture Notes in Physics}.

\parskip=7pt
\vfil \eject

\centerline{\bf 1995 Aspen Winter Conference  on  Gravitational
Waves and Their Detection}
\medskip
\centerline{Sydney Meshkov, Caltech}
\centerline{syd@ligo.caltech.edu}

     The Conference was held in Aspen, Colorado, Jan.22 - 28, 1995
with fifty-three participants. The conference had three aims, each of
which was achieved as detailed below. One aim was to hold an
organizational meeting of the LIGO users community during the
conference in order to formulate a users charter and organizational
structure. To this end, a nearly day long set of open meetings on
Wednesday, Jan. 25, discussed organizational structures and modes of
communication for the gravitational wave research community in the era
of LIGO, VIRGO and the other planned interferometric detectors as well
as a new generation of resonant mass detectors. The overarching goal
that emerged during the discussion was the need to define the means by
which the future gravitational wave research community might
communicate data between detectors for coordinated analysis, the modes
of collaboration in new research proposals for LIGO and other
detectors, and ways in which the scientific opportunities offered by
the new detectors might be maximized.

     A second aim was to involve physicists from other fields. To
implement this, the program started with a day of overviews by
acknowledged leaders in the field. In addition, Kip Thorne gave an
outstanding public lecture, entitled "Black Holes and Gravitational
Waves: The Dark Side of the Universe." About a quarter of the
participants were from other areas such as Elementary Particle
Physics, Quantum Optics, and Precision Frequency Measurement.

     The third aim was to bring together gravitational physicists with
varying perspectives on how to best detect gravitational waves. The
first day of overviews by Schutz, Cutler, Matzner, Thorne, Shoemaker,
Hamilton, and Bender, was followed by a session on Acoustic Detectors
at which Weber, Blair, Johnson and Pizzella discussed the history and
present status of the field. Brief descriptions of all of the present
and planned interferometric detectors were given by Sanders, Flaminio,
Ward, Kawabe, and Blair. The LIGO Users discussion described above
started with remarks by Sanders and Berley, followed by extensive
comments by many participants. An entire day was devoted to detailed
discussions of Interferometer Subsystems and Technologies, with every
interesting area covered by Shoemaker( he did yeoman duty because Rai
Weiss couldn't attend), Flaminio, Mizuno, Kawashima, Whitcomb, Shine,
Saulson, D. Robertson, and Moriwaki. A session on Ideas for Future
Detectors was a platform for the clever suggestions of Ruediger,
Drever, Kimble and Braginsky. This was followed by a series of talks
on Data Analysis and Observation Planning given by Schutz, Finn,
Nicholson, Compton and Allen, which discussed the strategy and
methodology involved. The final session considered Space Based
Gravitational Wave Detectors. Wahlquist and Jafry described how this
is done with American and European spacecraft. Stebbins, Newell and
Richman discussed their work on Active Vibration Isolation
Systems. The Conference summary, a tour de force, by David Shoemaker,
concluded a most enjoyable and exciting conference.

\end